\documentclass[fleqn,12pt,twoside]{article}
\usepackage[headings]{espcrc1}

\readRCS
$Id: espcrc1.tex,v 1.2 2004/02/24 11:22:11 spepping Exp $
\usepackage{graphicx}

\newcommand{\AmS}{{\protect\the\textfont2
  A\kern-.1667em\lower.5ex\hbox{M}\kern-.125emS}}
\def\gsim{\raise0.3ex\hbox{$>$\kern-0.75em\raise-1.1ex\hbox{$\sim$}}}
\def\lsim{\raise0.3ex\hbox{$<$\kern-0.75em\raise-1.1ex\hbox{$\sim$}}}

\title{Direct photons in 200 GeV $p+p$, $d$+Au, and Au+Au from PHENIX}

\author{S. Bathe\address[my_UCR]{University of California at
        Riverside}\footnote{Building 510 C, Brookhaven National
        Laboratory, Upton, NY11973-5000, USA} for the PHENIX
        Collaboration\footnote{For the full PHENIX Collaboration
        author list and acknowledgements, see Appendix
        ``Collaboration'' of this volume.  }}
       
\runtitle{Direct photons in $p+p$, $d$+Au, and Au+Au from PHENIX}
\runauthor{S. Bathe}

\begin{document}

\maketitle

\begin{abstract}
Direct photons were measured with the PHENIX experiment in $p+p$,
$d$+Au, and Au+Au at $\sqrt{s_{NN}}=200$ GeV.  To tackle the $p_T$
region below 5 GeV/$c$, direct photons were measured through their
internal conversion into $e^+e^-$ in Au+Au collisions.
\end{abstract}

\section{Introduction}

Direct photons are a unique probe of the hot and dense matter created
at RHIC: they allow access to the initial, thermalized state of the
collision.  Their measurement, however, is challenging.  One has to
cope with a large background from hadronic decays.

Direct photon measurements in $p+p$ constrain the hard-scattering
contribution in Au+Au.  A comparison to $d$+Au allows to quantify
contributions from initial-state effects. In heavy ion collisions,
thermal direct photons constrain the temperature of the collision
system in its hottest phase while hard direct photons serve as a
crucial baseline for the interpretation of the high-$p_T$ hadron
suppression observed earlier \cite{Adler:2003qi}.

\section{Direct Photon Analysis Via Low-Mass Electron Pairs}

The conventional direct photon measurement with the EMCal has been
described elsewhere \cite{Adler:2005ig,Okada:2005in}. In the following
we will focus on the measurement through internal conversion of direct
photons into $e^+e^-$.  Compared to the conventional measurement,
this technique improves both the signal-to-background ratio and the
energy resolution at intermediate $p_T$ where thermal production is
expected to contribute substantially \cite{Turbide:2003si}. The
measurement relies on a combination of excellent mass resolution at
low invariant mass, $m_{ee}$, and a low conversion probability due to
little material in the aperture. It has been carried out in heavy ion
experiments for the first time \cite{Cobb:1978gj,Albajar:1988iq}.

The full 2004 data set of about 900 M minimum bias events was
analyzed.  Events and centrality were selected as described in
\cite{Adcox:2003nr}.  Electrons in the central arms were identified by
matching charged particle tracks to clusters in the EMCal and to rings
in a ring imaging \v{C}erenkov (RICH) detector.

To illustrate the underlying idea we consider the $\pi^0$ Dalitz decay
where one decay photons is a virtual photon that further decays into
an $e^+$-$e^-$ pair. The invariant-mass distribution of the virtual
photon is given by \cite{Kroll:1955zu}

\begin{figure}[t]
\begin{minipage}[t]{0.525\linewidth}
\includegraphics[width=\linewidth]{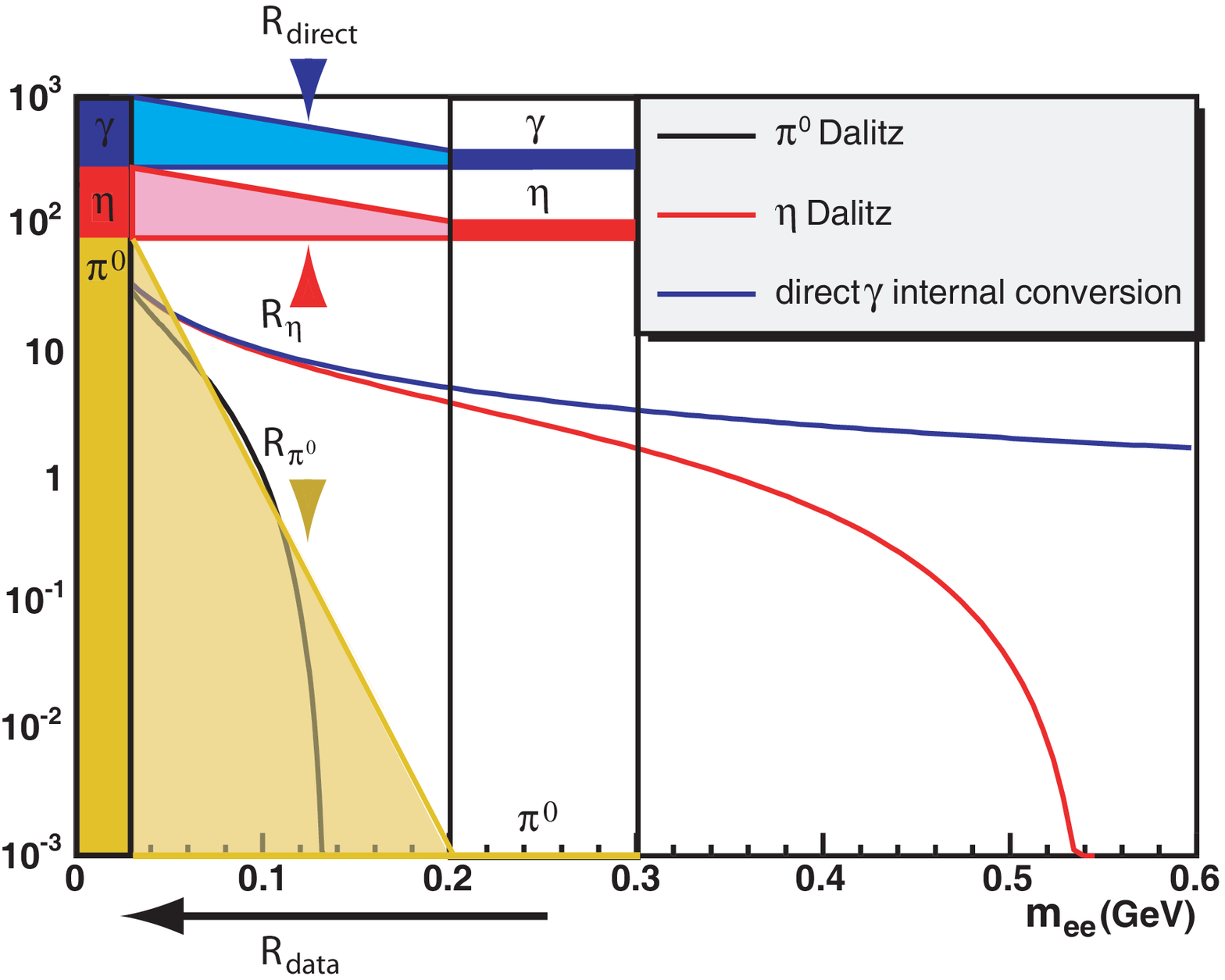}
\vspace*{-10mm}
\caption{\label{fig:fig1_sketch} Invariant-mass distribution of
virtual photons from the $\pi^0$ and $\eta$ Dalitz decays as well as
from direct photons according to Eq.  \ref{eq:Wa}.  It is illustrated
how the various contributions decrease to a fraction $R$ when going to
higher invariant mass with the $\pi^0$ contribution exhausting.}
\end{minipage}
\hspace{\fill}
\begin{minipage}[t]{0.425\linewidth}
\includegraphics[width=\linewidth]{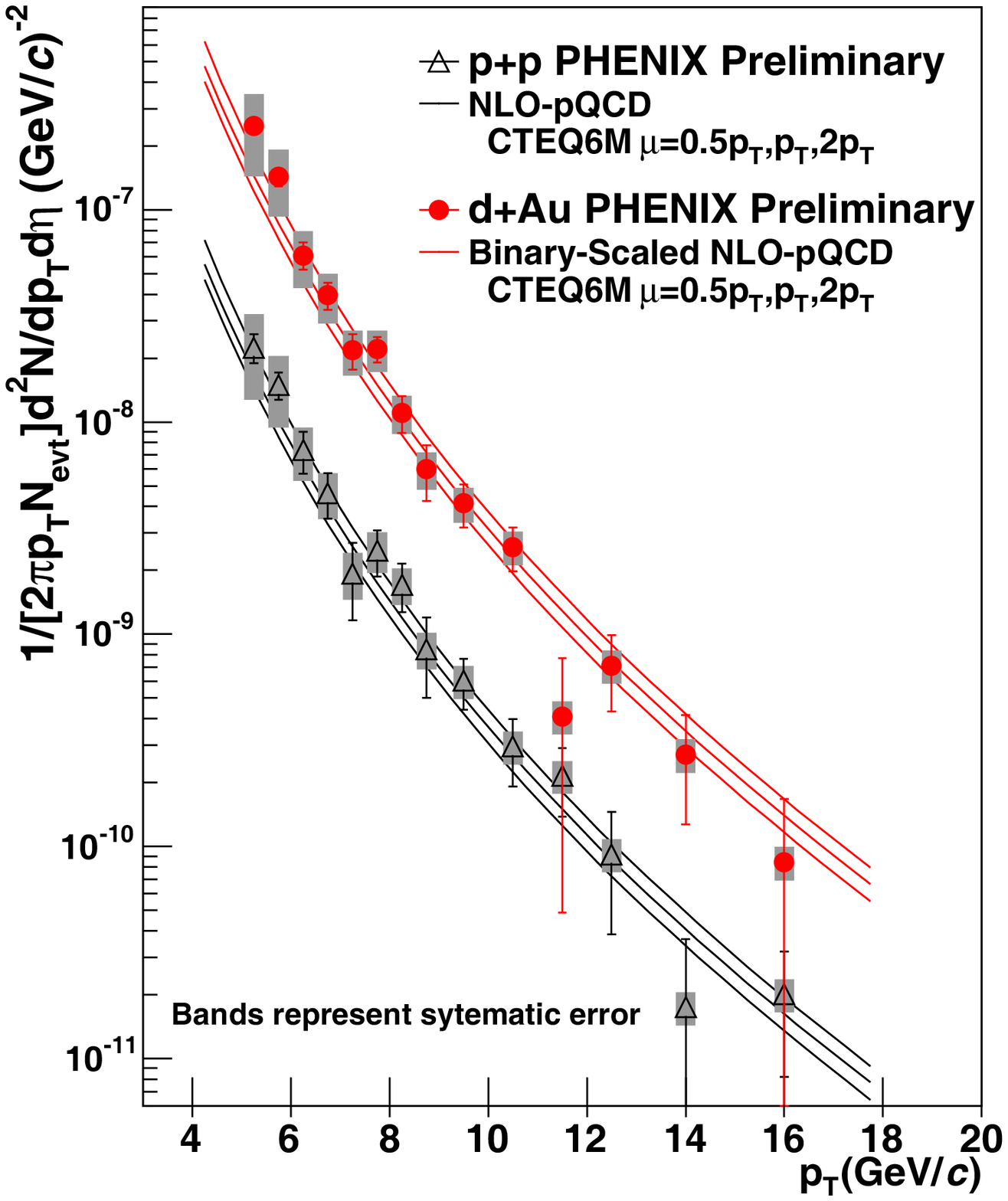}
\vspace*{-10mm}
\caption{\label{fig:fig2_ppdAu} Preliminary direct photon invariant
yield as a function of $p_T$ for $p+p$ \cite{Okada:2005in} and
minimum-bias $d$+Au collisions at $\sqrt{s_{_{NN}}}$ = 200~GeV.  The
solid curves are pQCD predictions \cite{Gordon:1993qc} for three
different scales that represent the uncertainty on the calculation.}
\end{minipage}
\end{figure}

\begin{equation}
  \frac{1}{N_\gamma} \frac{dN_{ee}}{dm_{ee}} = \frac{2\alpha}{3\pi}
   \sqrt{1 - \frac{4m_e^2}{m_{ee}^2}} (1 + \frac{2m_e^2}{m_{ee}^2}) 
   \frac{1}{m_{ee}} \mid F(m_{ee}^2) \mid^2 (1 - \frac{m_{ee}^2}{M^2})^3 \; ,
 \label{eq:Wa}
\end{equation}
and depicted in Fig. \ref{fig:fig1_sketch}.  In general, any source of
real photons produces virtual photons with very low invariant mass.
The rate and mass distribution of those virtual photons is given by
the same formula\footnote{Since the virtual photons decay in the
medium, this relation might be slightly modified.  This would not
affect the significance of the observed excess of direct photons, only
its translation into an absolute yield of real direct photons.}.  For
direct photons the phase space is not limited when $m_{ee} <<
p_T^{\mathrm photon}$.

To obtain a clean invariant-mass distribution of $e^+$-$e^-$ pairs,
pairs originating from photon conversions in the beam pipe or detector
material are rejected by an analysis cut.  The combinatorial
background is removed by the mixed-events technique. In this
measurement decay photons can mostly be eliminated by measuring the
yield of $e^+$-$e^-$ pairs in an invariant-mass region where pairs
from the $\pi^0$ Dalitz decay are largely suppressed due to their
limited phase space.  In order to convert the measured virtual photons
into real photons, the obtained yield has to be related to the yield
in an region where the phase space is unrestricted, so that
$\gamma_{\mathrm direct}^\star/\gamma_{\mathrm incl.}^\star =
\gamma_{\mathrm direct}/\gamma_{\mathrm incl.}$.  The term
$\gamma_{\mathrm direct}^\star/\gamma_{\mathrm incl.}^\star$ can be
calculated from the ratio of total yields in the two intervals,
$R_\mathrm{data} = N(\mbox{90-300 MeV})/N(\mbox{0-30 MeV})$, which is
known from the measurement, and the ratios, $R_i, i =
\{\gamma_{\mathrm{ direct}}, \pi^0, \eta, \mathrm{other} \}$, for the
various contributions, which can precisely be calculated from
Eq. \ref{eq:Wa} (cf. Fig.\ref{fig:fig1_sketch}).  Decay photons from a
cocktail of hadrons are considered here.  Finally, the yield of real
inclusive photons known from the EMCal measurement \cite{Adler:2005ig}
is needed to obtain the yield of real direct photons.

Since only the ratio of decay photon yields in the two invariant-mass
intervals is needed, the uncertainty on the $\eta$-to-$\pi^0$ ratio of
about 20 \% \cite{Adler:2004ta} is the main source of uncertainty,
translating into an uncertainty of 20 \% on the measured direct photon
yield.  Other sources are the EMCal-measured inclusive photon yield
(10 \%) and the $e^+$-$e^-$-pair acceptance (5~\%).  The total
systematic uncertainty is 25 \%.

\section{Results}

The preliminary direct photon spectra from the conventional
measurement for $p+p$ \cite{Okada:2005in} and minimum-bias $d$+Au at
$\sqrt{s_{_{NN}}}$ = 200 GeV is shown in Fig.
\ref{fig:fig2_ppdAu}. For the whole $p_T$ range of 5 to 16 GeV/$c$ the
observed yield is consistent with a next-to-leading-order
perturbative-QCD (NLO pQCD) calculation \cite{Gordon:1993qc} for both
$p+p$ and $d$+Au.  This supports the validity of the pQCD calculation
as a reference for hard direct photon production in Au+Au.  As shown
by PHENIX in the 2002 run \cite{Adler:2005ig}, at high $p_T$ direct
photon production is consistent with the pQCD expectation also in
Au+Au.  In the following we will focus on the region at intermediate
$p_T$.

\begin{figure}[t]
\begin{center}
\includegraphics[width=0.9\linewidth]{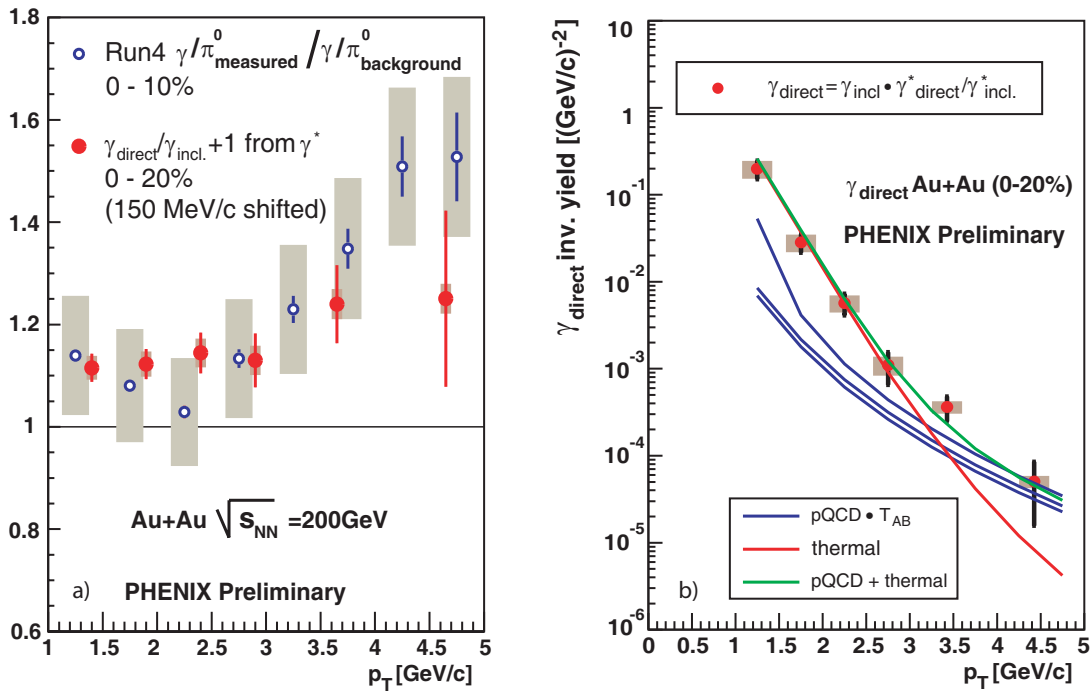}
\end{center}
\vspace*{-10mm}
\caption[]{a) Direct photon excess for the conventional and the
internal-conversion measurement in central Au+Au.  b) Direct photon
spectrum from the latter compared to pQCD \cite{Gordon:1993qc},
thermal-photon \cite{d'Enterria:2005vz}, and the sum of both
calculations.}
\label{fig:fig3_ratio_Aspe}
\end{figure}

In a preliminary analysis with the conventional method, a small but
stable subset of the 2004 data was selected to revisit the
intermediate-$p_T$ region, where no significant result had been
previously obtained for $p_T$ \lsim 3 GeV/$c$.  The new result is
shown in Fig. \ref{fig:fig3_ratio_Aspe} a) in terms of the double
ratio $(\gamma/\pi^0|_{\mathrm meas.})/(\gamma/\pi^0|_{\mathrm
bckgrd.})$.  This ratio indicates a direct photon excess as an
enhancement above 1.  There is still no significant excess below 3
GeV/$c$.  Further work is ongoing to increase the significance.

The preliminary result from the measurement of virtual photons is
presented in the same figure in terms of $\gamma_{\mathrm
direct}/\gamma_{\mathrm incl.} + 1$, which also indicates a direct
photon excess as an enhancement above 1.  (Note that the two
quantities aren't exactly equivalent.).  The result shows a
significant direct photon excess of about 10 \% for $1 < p_T < $ 5
GeV/$c$ and the 20 \% most central Au+Au collisions.  It is consistent
with the EMCal measurement.  In Fig. \ref{fig:fig3_ratio_Aspe} b) the
direct photon invariant yield from the virtual photon measurement is
shown and compared to various theoretical calculations.  With large
significance, a direct photon spectrum was obtained for $1 < p_T < $ 5
GeV/$c$.  The spectrum lies significantly above a $T_{AB}$-scaled pQCD
calculation \cite{Gordon:1993qc} for $p_T$ \lsim 3 GeV/$c$.  A 2+1
hydrodynamical model \cite{d'Enterria:2005vz} for thermal-photon
emission with an average initial temperature of $T_0^{\mathrm ave}=$
360 MeV ($T_0^{\mathrm max}=$ 570 MeV) and a formation time of
$\tau_0=$ 0.15 fm/$c$ underpredicts the data for $p_T$ \gsim 3
GeV/$c$.  The data can be described when both sources are combined.
The obtained temperature is only meaningful if the observed excess is
of thermal origin.  To confirm the result, an analysis of $p+p$ and
$d$+Au using the same technique is needed.  If the excess in Au+Au is
mainly from thermal photons the reference data will show a much
smaller effect.

\section{Summary}

Direct photons were measured with the PHENIX experiment in $p+p$,
$d$+Au, and Au+Au at $\sqrt{s_{NN}}=200$ GeV.  At high $p_T$ (\gsim 5
GeV/$c$), the measured yield in all systems is consistent with a NLO
pQCD calculation.  To tackle the $p_T$ region below 5 GeV/$c$, direct
photons were measured through their internal conversion into
$e^+e^-$ in Au+Au collisions.  With this powerful technique a
significant measurement for $1 < p_T < 5 \mbox{GeV}/c$ was achieved,
lying significantly above the NLO pQCD expectation, but consistent
with calculations when thermal photon emission is taken into account.

\bibliographystyle{unsrt.bst}

\bibliography{QM05PhotonsBathe}

\end{document}